\documentclass[a4paper,english,twocolumn,rmp]{revtex4}
\usepackage[T1]{fontenc}
\usepackage[latin9]{inputenc}
\usepackage{amsthm}
\usepackage{amsmath}
\usepackage{graphicx}
\usepackage{esint}

\makeatletter

\pdfpageheight\paperheight
\pdfpagewidth\paperwidth

\@ifundefined{textcolor}{}
{%
 \definecolor{BLACK}{gray}{0}
 \definecolor{WHITE}{gray}{1}
 \definecolor{RED}{rgb}{1,0,0}
 \definecolor{GREEN}{rgb}{0,1,0}
 \definecolor{BLUE}{rgb}{0,0,1}
 \definecolor{CYAN}{cmyk}{1,0,0,0}
 \definecolor{MAGENTA}{cmyk}{0,1,0,0}
 \definecolor{YELLOW}{cmyk}{0,0,1,0}
}
\numberwithin{equation}{section}
\numberwithin{figure}{section}

\makeatother

\usepackage{babel}
\begin{document}

\title{The remarkable discreteness of being}

\author{Bahram Houchmandzadeh}

\affiliation{Univ. Grenoble 1 / CNRS, LIPhy UMR 5588, Grenoble, F-38041, France}

\email{bahram.houchmandzadeh@ujf-grenoble.fr}

\begin{abstract}
Life is a discrete, stochastic phenomenon : for a biological organism,
the time of the two most important events of its life (reproduction
and death) is random and these events change the number of individuals
of the species by single units. These facts can have surprising, counter-intuitive
consequences. I review here three examples where these facts play,
or could play, important roles : the spatial distribution of species,
the structuring of biodiversity and the (Darwinian) evolution of altruistic
behavior. 
\end{abstract}
\maketitle

\section{Introduction.}

Many quantities in the physical world are continuous and measured
by real numbers: positions, speeds, concentrations, weights, etc.
In many areas of science, however, it was realized that complex patterns
can be explained by supposing the existence of discrete underlying
levels that can be described using integers. In chemistry, the various
laws of composition of elements such as ``definite proportions''
and ``multiple proportions'' known around 1800 AD led Dalton to
formulate the atomistic theory and give a simple, elegant explanation
to all these laws. Around 1900 AD, Planck, Einstein, Bohr and others
realized that the most daunting problems of the (then) modern physics
such as the radiation spectrum of stars, the universal temperature
dependence of the specific heat of solids, the speed of electrons
ejected by solids under radiation (photoelectric effect), ... can
be solved elegantly by supposing that the energy (or action) is quantified
and varies only in integer units. In biology, the theory of Darwinian
evolution was inconsistent with the then obvious blending theory of
inheritance. The work of Mendel, and its rediscovery by de Vries,
Correns and Tschermak around 1900 AD (curious coincidence) restored
the mathematical consistency of Evolution by introducing the concept
of genes as the quantum of inheritance information (see the first
chapter of Fisher's book\cite{Fisher1999}). 

These are but a few examples where complex patterns could be simply
explained by supposing an underlying discrete level. The discreteness
hypothesis, and specially its consequences, was in each of these cases
unintuitive. Living organisms on the other hand do not need the discreteness\emph{
hypothesis}, as this is the most obvious fact about them : death and
birth events change their number by integers only. This obvious fact
can have non-intuitive consequences\cite{Durrett1994}, and we will
review three such cases : (i)  spatial clustering of organisms, which
is observed for nearly all living organisms ; (ii) the observed biodiversity
and many of its general laws such as the species-area relationship
; (iii) the emergence of cooperative behavior during  Darwinian evolution.
In all these cases, the reason for the importance of discreteness
in living organisms is related to the unpredictability (stochasticity)
of many phenomenon of life. For example, the moment of death or duplication,
the direction of movement for food search, ... are more or less random
for an individual. One of the most fundamental result of stochastic
phenomena is that the unpredictability of a system as a whole diminishes
as the number of its constituent grows. When individuals can be counted
by integers, the number of the constituents is finite and stochasticity
can be an important player, inducing counter-intuitive effects. The
prime examples of this phenomenon are chemical reactions. In a typical
test tube reaction, there are $\sim10^{23}$ molecules, a number so
large that the unpredictability of molecules colliding to form bounds
is erased and chemical reactions are perfectly predictable. In a bacteria
however, some molecules are present at such a low level as to be countable
and chemical reactions associated to these molecules can become extremely
noisy. This phenomenon, first noticed by Novick and Weiner for the
transcription of Lac operon \cite{Novick1957} has given rise to the
field of epigenetics and non-genetic individuality\cite{Houchmandzadeh2011a}. 

An important remark is in order. In none of the cases reviewed here
is it claimed that a simple discrete theory will explain all the phenomena.
I only observe that discreteness implies some surprising patterns
which always exist. When complex patterns are observed in nature,
the contribution of discreteness should be subtracted and only the
remaining part, if any, needs a special theory.

\section{Spatial Clustering.}

Since the 1970's and the gathering of large amount of data on spatial
distribution of various species, ranging from plants to insects to
mammals, it has become obvious that nearly all species tend to have
a clustered distribution and to aggregate into some areas\cite{Taylor1978}.
The study of these spatial distributions has now become an independent
field and is called metapopulation biology or ecology (for a review,
see \cite{Hanski1997,Hanski2004}). If the diffusion of organisms
(animals move and plants disperse their seed) were random, one would
expect that the distribution of species would soon (in few generations)
become homogeneous (everything is everywhere\cite{O'Malley2007}).
This is analogous for example to the dilution of a drop of ink in
water. Common sense therefore requires that if we observe aggregation
of individuals in one place we should look for deterministic causes.
There is no shortage of deterministic causes : (i) species are adapted
to some environment,  nature is heterogeneous therefore each species
tends to concentrate in places (sub-habitats) to which it is best
adapted ; (ii) many species are social and their social interaction
could be a cause of aggregation. These are two of the most studied
explanations of clustering of organisms.

\begin{figure}
\begin{centering}
\includegraphics[width=0.95\columnwidth]{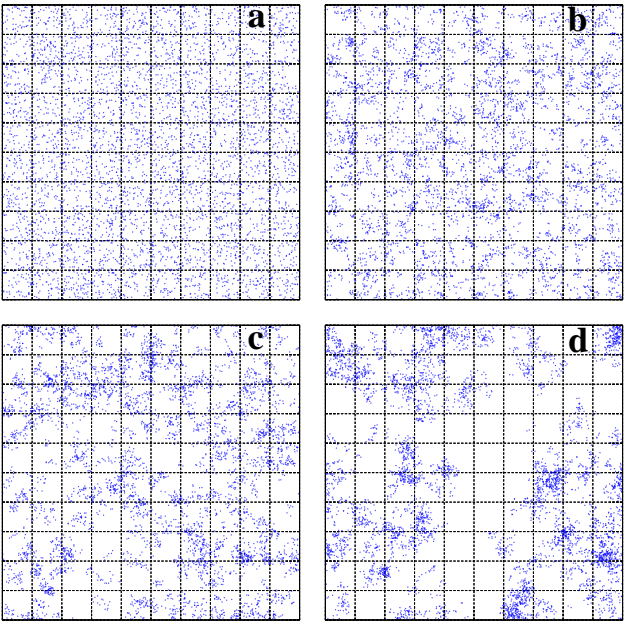}
\par\end{centering}

\caption{Spatial distribution of 5000 individuals (dots) with increasing patchiness.
The VMR computed over the displayed grid is (a) $1$ (homogeneous)
; (b) $5.1$; (c) 13.8 and (d) 49.1.\label{fig:distributions} }

\end{figure}

Plain common sense however is wrong in this case: the sole fact of
 discreteness of life is enough to cause clustering and no amount
of random movement can counteract this agglomeration. This is what
we  show below. Before going further however, we should define precisely
what we mean by clustered distribution and how we can measure it.
The most practical way of measuring patchiness is to divide  space
into squares (quadrats), count the number $n_{i}$ of individuals
in each square, compute the mean $\left\langle n\right\rangle $ and
the variance $V$ of these numbers, and calculate the variance to
mean ratio (VMR) $V/\left\langle n\right\rangle $. (Figure \ref{fig:distributions})
. For homogeneous random distributions the VMR is equal to unity (Fig
\ref{fig:distributions}a); for clustered distributions VMR>1. VMR<1
denotes regularities (such as in human plantations) and is seldom
encountered in Nature. The VMR is a robust measure of patchiness;
it is a function of  the grid size and this size dependence can be
used as a test of the underlying theory causing the clustering. This
measure can also be related to other popular metrics such as box counting
and fractal dimension measurements.

Let us come back to the problem of clustering. Consider a collection
of simple and similar organisms all moving randomly with diffusion
coefficient $D$, reproducing at rate $\alpha$ and dying at rate
$\mu$. A naive model of the distribution of these ``Brownian bugs''
would use a diffusion equation for their concentration $c(x,t)$ of
the form of 
\begin{equation}
\partial_{t}c=D\Delta c+(\alpha-\mu)c\label{eq:diffusion}
\end{equation}
The diffusion equation used by Fick to model concentration propagation
is widely used in ecology (\cite{Okubo2002}. It is a consequence
of ``particles flow from high to law concentrations''. The left
hand term denotes changes (per unit of time) in the concentration.
In the right hand, the first term ($\Delta c$ ) is the balance of
particle movements at one position due to flow from neighboring positions
; the second term states that the number of birth and death at one
position is proportional to the number of individual present at this
position.

Consider for example the particular case where birth and death rates
are equal. Then we will have a plain diffusion equation and any spatial
heterogeneity will be smoothed out after some time, as for the example
of the drop of ink in water. Young et al \cite{Young2001} used such
a simple model of Brownian bugs to study the phenomenon of plankton
blooms, but instead of resorting to equation (\ref{eq:diffusion}),
they numerically simulated these bugs and found the exact contrary
of the expected phenomena : the distribution, which was homogeneous
at the initial time, would get more and more patchy as time passes.
Something is grossly wrong with the use of the continuous approach
(\ref{eq:diffusion}) : such equations are written for averaged quantities
and suppose that fluctuations ( deviations from the mean) are small
compared to the mean. However, the noise of reproduction and death
due to the discreteness of living organism violates this assumptions:
as we will see below, fluctuations become much larger than the averages,
hence the error in using continuous differential equations in modeling
ecological systems.

\begin{figure}
\begin{centering}
\includegraphics[width=0.8\columnwidth]{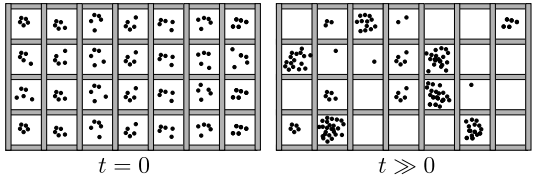}
\par\end{centering}

\caption{Reproduction/death noise : placing $n_{0}$ individuals in each cell
at time $t=0$ and letting them die and reproduce at the same rate.
As time passes, some cells will contain a large number of individuals,
while others become empty. \label{fig:0dclustering}}

\end{figure}

To understand reproduction/death induced fluctuations, we need to
include the second important aspect of life : for an individual, the
moment of its death or reproduction is a random variable. For a collection
of $n$ individuals, we can speak only about the probability of a
death/birth occurring during a given time interval. If the probabilities
induce small fluctuations, the stochastic process can be approximated
by a differential equation (mean field approach, see appendix) ; if
not, one has to resort to the Master equation approach (and/or its
numerical resolution) in order to estimate various statistical quantities
such as the mean, the variance and the correlations (see appendix).
Let us forget about the spatial aspect of the problem at hand for
the moment. Consider the space divided into non-communicating habitats
and let us place exactly $n_{0}$ individuals in each cell at time
$t_{0}=0$ (Figure \ref{fig:0dclustering}). These individuals are
capable only of reproducing/dying. Let $n(t)$ be the number of individuals
in a cell at time $t$. In the simplest possible model when birth
at rate $\alpha$ and death at rate $\mu$ are constant and independent
of density, age structure, previous births ..., the probability density
for one birth/death to occur is proportional to the number of of individuals
$n$ and independent of time. It reads 
 
\begin{equation}
W^{+}(n)=\alpha n\,\,\,;\,\,\, W^{-}(n)=\mu n\label{eq:transition rates}
\end{equation}
and from the above transition rates, the probability $P(n,t)$ of
observing $n$ individuals at time $t$ is deduced through the book
keeping (Master) equation (see appendix). We can very easily numerically
simulate the above stochastic process and observe that as time increases,
many cells will become empty and a few will harbor a very large number
of individuals. The average number of individuals per cell will remain
constant, but the variance and hence the VMR will increase linearly
as a function of time. This is a system which, after a few generation,
will display a huge amount of clustering. In fact, in this simple
stochastic process, the time evolution of the mean $\left\langle n\right\rangle $
and the variance $V=\left\langle n^{2}\right\rangle -\left\langle n\right\rangle ^{2}$
of the number of individuals in cells can be deduced exactly from
the Master equation. For the case where $\alpha=\mu$ and $n_{0}$
individuals per cell at $t=0$:
\[
\left\langle n(t)\right\rangle =n_{0}\,\,\,;\,\,\, V(t)=2n_{0}t
\]
Therefore, even though the average number of individuals per cell
remain constant, the variance increases as a function of time, a behavior
pictured in Figure(\ref{fig:0dclustering}) \cite{Houchmandzadeh2002}. 

What is special about this reproductive noise is the fact that it
cannot be canceled out by diffusion. Removing the barriers between
the cells and letting the individuals diffuse from high density to
low density sites only slows down the clustering phenomenon, but does
not inhibit it \cite{Houchmandzadeh2002} and the VMR still increases
with time for one- and two-dimensional ecosystems. In the case $\alpha=\mu$:
\begin{eqnarray*}
\mbox{VMR} & \sim & \sqrt{t}\,\,\,\,\mbox{for }d=1\\
 & \sim & \log t\,\,\,\mbox{for\,}d=2
\end{eqnarray*}
The two dimensional case corresponds exactly to what Young et al.
\cite{Young2001} were observing in their numerical simulations. Note
that the prediction of a pure diffusion equation is that the VMR will
stay at the value 1 at all time. 
\begin{figure}
\begin{centering}
\includegraphics[width=0.9\columnwidth]{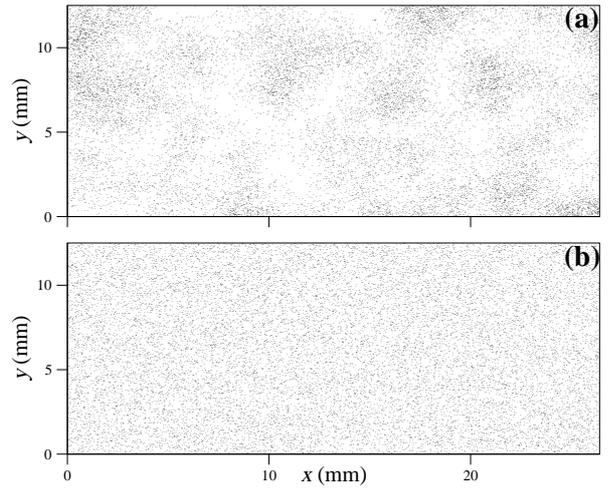}
\par\end{centering}

\caption{Neutral clustering of microorganisms : (a) \textasciitilde{}70000
Dictyostelium Discoïdum (size \textasciitilde{}10 $\mu$m) after approximately
7 generations. Each dot represents a single individual and its position
was measured from an actual experiment \cite{Houchmandzadeh2008};
the VMR$\approx40$ for squares of size $0.7$ mm ; (b) the same number
of individuals spread homogeneously.\label{fig:clusteringexp}}

\end{figure}

So all living organisms will naturally form spatial clusters, at least
in the simplest, neutral models. Can this clustering be observed experimentally?
The answer is yes. Spread some microorganism capable of movement,
reproduction and death on a Petri dish, measure the position of \emph{each}
one at each time step and compare it to the predicted auto-correlation
function or VMR (figure \ref{fig:clusteringexp}). In such a controlled
experiment, all the parameters ($\alpha,\mu,D$) are measured and
there is no room for free fitting parameters. The only difficult step
is to measure the position of all microorganisms, which can be achieved
by an automatized microscope and image analysis. The experiment which
was indeed performed \cite{Houchmandzadeh2008,Houchmandzadeh2009a},
showed the perfect agreement of the spatial autocorrelation function
with the theoretical computations.

Real ecosystems are density dependent and the density of individuals
cannot grow very large. We can incorporate this density dependence
into the model in its most stringent case : the ecosystem is composed
of many species, the total density is fixed, and when one individual
dies, it is replaced by the progeny of a neighboring one, whatever
its species. In the framework of neutral ecosystems, \emph{i.e. }all
individuals similar in their birth, death and dispersal properties,
it can be shown that the major features exhibited above do not change
and individuals still form increasingly large clusters, uniform in
their composition of species \cite{Houchmandzadeh2003}. This clustering
can also be (painfully) measured in real ecosystems (for example in
rain forests\cite{Condit2002}) and shown to be compatible with the
theoretical computations, although in real ecosystems, many parameters
cannot be measured. 

Spatial clustering of organisms is one of the most fundamental problems
in ecological studies. The message of this section is the following
: observing a patchy distribution should not be considered per se
as surprising and one should not rush to find deterministic causes
for it. The very nature and discreteness of life naturally leads to
clustering. Of course, all clustering are not caused by discrete effects.
Before looking for other causes however, one must subtract the effect
of neutral causes and use deterministic causes only for the remaining
(if any) patchiness.

\section{Neutral biodiversity.}

Observation of the stunning biodiversity in various ecosystems is
one of the factors that led Darwin and Wallace to formulate the theory
of evolution. The finches of Galapagos are the standard example cited
in any textbook of the field\cite{Lo}. Even at a single trophic level,
\emph{i.e. }considering species which use the same resources, the
biodiversity is always large. In spite of many competing theories
the question of the causes of  biodiversity is still unanswered today.
The adaptationist program criticized by Gould and Lewontin \cite{gould1979}
is still predominant: each species is adapted to its local environment
and biodiversity is just a reflection of the heterogeneity of Nature.
Neutral biodiversity can exist only because of geographical barriers
between close ecotypes. The possibility of having speciation at the
same trophic level at the same geographical location has been ruled
out by Ernst Mayr in his famous book\cite{Mayr1942}, with far-reaching
consequences on evolutionary thinking.

Ecologists however began to gather large data on biodiversity and
observed general patterns everywhere. One of the most striking observed
``law'' is the species-area relationship which states that the number
of species $S$ in an area exhibits a power law dependence on  the
size $A$ of the area considered: $S=kA^{z}$ with $z$ in the {[}0.2,0.3{]}
range for most habitats\cite{Williamson2008}. An alternative and
more precise measure of biodiversity for a fixed area is the abundance
curve: collecting species in a given area and measuring the abundance
of each species leads to the abundance curve $\phi(n)$, which is
the histogram of the number of species having abundance $n$. Abundance
curves taken from very different habitats began to show very similar
patterns (for a review, see \cite{Hubbel2001}). The third observation
came from measurements of biodiversity in islands close to a continent.
It was observed that the number of species in islands decreased as
a function of its distance from the continent and increased with the
size of the island. 

To explain the third observation, MacArthur and Wilson\cite{MacArthur1963,MacArthur1967}
took a bold approach. They supposed that (i) all species at the same
trophic level are \emph{equivalent} ; (ii) species migrate from continent
to islands, with the rate of migration a decreasing function of the
distance ; (iii) due to random sampling from one generation to the
other, species become extinct in islands, with the extinction rate
a decreasing function of the size of the island. The number of species
present on the island is then a dynamic equilibrium between migration
and extinction. 

MacArthur and Wilson's article, considered as a cornerstone of biogeography,
was a radical departure from Mayr and the adaptationist program, and
proved extremely successful. The next radical step then was taken
by Hubbell\cite{Hubbel2001} who applied the same idea to the whole
continent : all species at a given trophic level are equivalent, new
species appear by mutation and become extinct by genetic drift. The
biodiversity curve is then a function of a single number that takes
into account the mutation rate and the size of the community. Hubbell's
book founded what is called the neutral theory of biodiversity and
provoked an incredibly wide and heated debate in the ecological community,
which is still ongoing. 

I review below some of the mathematical consequences of the neutral
theory to which we contributed. In retrospect, it seems strange that
the idea of neutrality, considered very early by population geneticists
such as Malecot\cite{Malecot1948} and Kimura\cite{Kimura1985}, took
so much time to permeate the ecological/evolutionary thinking; I believe
that this is partly due to the influence of Mayr's book\cite{Butlin2008}
and the prevalence of the competitive exclusion principle\cite{Hardin1960}.
The main idea however is very simple : neutral macroecology is similar
to neutral population genetics\cite{Hu2006}. The latter deals with
alleles of a gene, their frequency and its change because of genetic
drift and (neutral) mutations, where the former deals with the equivalent
concepts of species, their abundance and its change because of ecological
drift and neutral speciation (presumably because of the accumulation
of many mutations at the individual levels) and so on. New species
emerge with a rate $\nu$ . It takes some times $\tau_{a}$ for a
new species to become abundant by pure genetic drift. If the arrival
time of new species $\tau_{e}$ is much shorter than $\tau_{a}$,
many equivalent species will coexist at the same geographical location
and their abundance will be a dynamic interplay between emergence
of new species and extinction of existing one. 

\begin{figure}
\begin{centering}
\includegraphics[width=0.9\columnwidth]{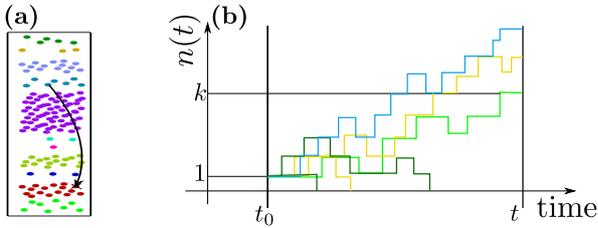}
\par\end{centering}

\caption{(a) The Moran model of a neutral community composed of various species
(distinguished here by their colors), where an individual is replaced
upon its death by the progeny of another regardless of its species.
(b) Each new species appears with abundance 1 by mutation at some
time $t_{0}$; Stochastic dynamics of the number of individuals $n(t)$
of few species appeared at time $t_{0}$. \label{fig:neutralspeciation}
.}
\end{figure}

Consider a community consisting of $N$ individuals and $S$ species,
with species $i$ having $n_{i}$ individuals (Figure \ref{fig:neutralspeciation}a).
 All individuals, regardless of their species, are equivalent in their
reproductive/death rate. When an individual dies, it is immediately
replaced by the progeny of another one. Because of mutations, the
progeny can differ from its parent with probability $\nu$, thus forming
a new species appearing with abundance 1. After its appearance, the
species abundance is a stochastic function of time ; if an individual
is the sole representative of a species and dies, then this species
disappears. As in the previous section, the probability $P(n,t|1,t_{0})$
for species $i$ to have $n$ individuals at time $t$, knowing the
species appeared at time $t_{0}$, obeys a Master equation where the
transition rates are\cite{Vallade2003a} (see appendix ): 
\begin{eqnarray}
W^{+}(n) & = & \mu(N-n)n(1-\nu)/N\label{eq:metacom1}\\
W^{-}(n) & = & \mu n\left(N-n+\nu(n-1)\right)/N\label{eq:metacom2}
\end{eqnarray}
The increase rate $W^{+}(n)$ is the probability density of death
of an individual that does not belong to the considered species $\mu(N-n)$
multiplied by the probability of birth of an individual that belongs
to the considered species $n/N$, times the probability of no mutation
$(1-\nu)$. The decrease rate is similar, but takes also into account
the probability of an individual dying and being replaced by the progeny
of a member of its own species with a mutation. 

Let us set the origin of time at $t_{0}=0$. The master equation gives
the fate of one particular species. $\left\langle \phi(n)\right\rangle $,
the average number of species having population size $n$ at time
$t$ is the sum of all those who have been generated at an earlier
time $\tau$ and have reached abundance $n$ at time $t$ : 
\begin{eqnarray*}
\left\langle \phi(n)\right\rangle  & = & \int_{0}^{t}f(\tau)P(n,t|1,\tau)d\tau\\
 & = & \nu\int_{0}^{t}P(n,t-\tau|1,0)d\tau\\
 & = & \nu\int_{0}^{t}P(n,\tau|1,0)d\tau
\end{eqnarray*}
where $f(\tau)$ is the probability per unit of time of generating
a mutant and is equal to $\nu$ (time is measured in units of generations
$1/\mu$). Defining the mutation pressure as $\theta=N\nu$, the quantity
$\phi$ can be obtained at the limit of large times and shows that
an equilibrium is reached. For large communities, using proportions
$\omega=n/N$ and abundances $g(\omega)=N\left\langle \phi(n)\right\rangle $,
the result takes a simple form\cite{Vallade2003a} 
\begin{equation}
g(\omega)=\theta\omega^{-1}(1-\omega)^{\theta-1}\label{eq:metaabund}
\end{equation}
The above computations ignore spatial distances: an individual can
be replaced only by the progeny of its neighbor rather than by everyone
in the community. A self consistent model of geographical dispersal
is incredibly difficult. We can however go one step further and apply
the above model to the case of island biogeography, where a small
island of size $M$ is close to a continent of size $N$ ($M\ll N$).
The population of the island is affected by migration from the continent,
but given the large size of the continent, the reverse is not true.
We can also neglect mutation inside the island as the mutation pressure
is small. So the transition rates in the island are similar to eqs
(\ref{eq:metacom1},\ref{eq:metacom2}) except that a local individual
can be replaced by a migrant from the continent with probability $m$,
where the abundances are given by expression (\ref{eq:metaabund}).
Defining the migration pressure as $\xi=Mm$, in the limit of large
sizes of both the island and the continent, we can compute the relative
abundance $g_{I}(\omega)$ inside the island as\cite{Vallade2003a}
\begin{equation}
g_{I}(\omega)=\xi\theta\int_{0}^{1}(1-\omega)^{\xi u-1}\omega^{\xi(1-u)-1}u^{\theta}du\label{eq:gw}
\end{equation}
This expression may seem cumbersome, but it can be easily plotted
and depends on only  two parameters : $\theta$ which itself can be
seen as a function of biodiversity on the continent and $\xi$ which
is a simple decreasing function of the distance between the continent
and the island. This expression was also obtained by Volkov et al.\cite{Volkov2003}
and in a slightly modified form by Etienne\cite{Etienne2005}. Expression
(\ref{eq:gw}) is the mathematical expression of the original MacArthur
and Wilson model and can be put to experimental verification.

Improving the above model by taking fully into account the spatial
dimension seems mathematically intractable. We have been able to slightly
improve the continent-island model by treating both communities on
an equal footing\cite{Vallade2006} but going further seems beyond
the reach of the mathematical tools we  used. Nevertheless, the neutral
theory of biodiversity is a falsifiable theory of biodiversity. It
has been put to intense test and has been proved  successful at interpreting
quantitatively available data in island biogeography\cite{Rosindell2011}.
As in the previous section, the merit of this model is to provide
a first approximation for biodiversity which will always be present,
even though many data will necessitate the addition of more ingredients,
such as for example, density dependence of replacement rates\cite{Jabot2011,McGill2006}
to explain deviation from this theory.

\section{Emergence of altruistic behavior in Darwinian evolution.}

Altruistic behavior is widespread among living organisms. ``Altruism''
is an emotionally charged term that many scientists avoid in favor
of more neutral terms such as cooperative behavior. We stick to this
word here and define altruistic behavior as the production of some
``common good'' that benefits all individuals of the same species
in the community, at a cost to the producer. From the evolutionary
point of view, fitness is the mean number of surviving and reproducing
descendant of individuals of a given genotype. The cost and benefit
here are all measured in units of fitness.

Light production in \emph{Vibrio fischeri}\cite{Visick2006,Foster2004}\emph{,}
siderophore production in \emph{Pseudomonas aeruginosa}\cite{West2003},
invertase enzyme production in \emph{Saccharomyces cerevisiae}\cite{Gore2009},
stalk formation by \emph{Dictyostelium discoideum}\cite{Kessin2001,Foster2004},
are but a few examples, taken from the microbial world, where individuals
in a community help others at their own cost by devoting part of their
resources to this help. From the evolutionary point of view, altruists
have a lower fitness than other individuals in the community who do
not help, but are recipients of the benefits produced by altruists.
Throughout this paper, we call these latter individuals `selfish'. 

How can altruistic behavior  emerge by natural selection if individuals
having a genotype which displays this trait have a lower fitness than
the selfish ones ? This is among the hottest debates of  evolutionary
biology, and has been ongoing from the inception of the discipline\cite{Dugatkin2006}.
In the deterministic view of evolution, genotypes with higher fitness
increase their frequency in the population ; therefore, if altruism
is selected it means that its associated genotype has some hidden
benefits that compensate its apparent lower fitness. The only task
is to discover the hidden advantage.

The first class of model for the hidden advantage was proposed by
Hamilton\cite{Hamilton1964a,Hamilton1964b} and is known as kin selection:
the common good is not provided to everybody, but only to individuals
related by common descent (kins). The original Hamilton model is based
on ``frequency dependent fitness'' and was formulated for sexually
reproducing organisms. This model and other mathematically equivalent
models such as direct or indirect reciprocity\cite{Nowak2006} can
be easily understood as follow for asexual organisms. Consider the
simplest case where an allele of gene confer a fitness $r$. The deterministic
Fisher equation for the change in the frequency (relative abundance)
$p$ of this allele is \cite{Ewens2004}:
\begin{equation}
dp/dt=(r-1)p(1-p)\label{eq:detrm}
\end{equation}
This equation is derived from the original Fisher-Wright model of
population genetics : in a community of size $N$ (where $N$ is supposed
to be large), each individual produces progeny proportional to its
fitness and $N$ individuals are selected randomly among all the progeny
to constitute the next generation. 

Advantageous mutants have fitness $r>1$ and therefore increase their
frequency, where deleterious mutants have fitness $r<1$ and decrease
their frequency. This equation supposes that fitness does not depend
on the frequency of the allele and is a constant
\begin{figure}
\begin{centering}
\includegraphics[width=0.85\columnwidth]{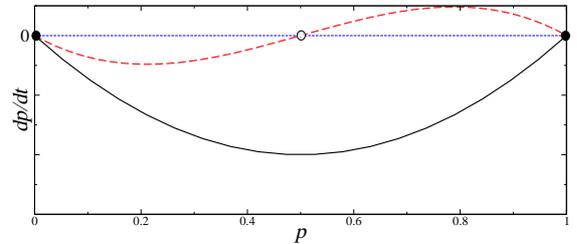}
\par\end{centering}

\caption{the phase space presentation of the Fisher equation (\ref{eq:detrm}).
Solid curve: constant fitness $r-1<0$ ; of the two fixed point, only
$p=0$ is stable. Dashed curve: frequency dependent fitness where
$r-1=f(p)<0$ for $p<p^{*}$. Both fixed points $p=0$ and $p=1$
are stable.\label{fig:fixedpoint}}

\end{figure}
. If however the excess fitness $r-1$ is a function of the gene frequency
$f(p)$ and the function changes its sign for some intermediate frequency
$p^{*}$, then the gene will increase its frequency if $p>p^{*}$
(figure \ref{fig:fixedpoint}). This is precisely the point made by
Hamilton : if help is provided and received only among altruists,
then at high frequency, the benefits that each altruists receive from
other `kins' or other altruists can outweigh the cost of the common
good production to one individual. Nowadays these models are mostly
studied through the game theory language and the frequency dependent
fitness $f(p)$ is obtained through a payoff matrix ; the frequency
dependent equation for the rate of change is called replicator dynamics\cite{Nowak}. 

The second class of models, called group selection, suppose that individuals
are divided into groups. Not only do individuals compete inside each
group in order to increase their frequency, but groups compete among
each other at a higher level of selection\cite{Traulsen2005}. The
idea of group selection goes back to the inception of evolutionary
biology and was promoted by the founding fathers of modern evolutionary
synthesis, then was discredited by G.C. Williams\cite{Williams1966}
, then restored by Lewontin\cite{Lewontin1970} and Price\cite{Price1970}
and regained respectability again in the 1990's. 

This two class of models are nowadays the main explanations for the
emergence of altruisms in Darwinian evolution\cite{Nowak2006}, even
though a ``religious'' war can erupt between them from time to time
(see for example\cite{Nowak2010} and some among many replies to it\cite{Boomsma2011,Strassmann2011}).

As in the two previous section, I want to review an alternative theory
we developed\cite{Houchmandzadeh2012a}, based on the discreteness
and the genetic drift it causes. As in the previous sections, I do
not claim that this theory explains all the observed behavior and
replaces the other two. But as in the previous section, I show that
a very simple explanation exists which does not rely on some hidden
benefits. This simple model applies to cases where the production
of common good or the cooperative behavior increases the carrying
capacity of the habitat.

The deterministic Fisher equation (\ref{eq:detrm}) is not satisfactory
at small excess relative fitness: a fitter genotype appearing at one
copy number can disappear just by chance and not get the possibility
of increasing its frequency at all. Evolutionary dynamics is a stochastic
process due to competition between deterministic selection pressure
and the inevitable role of chance factors in influencing who reproduces,
how many children they have, and so on. In order to capture the main
characteristics of this competition, Fisher and Wright introduced
a very simple model which was later slightly modified by Moran(\cite{Moran1962})
to make it mathematically more tractable\cite{Houchmandzadeh2010}.
The model consists of a community of fixed size $N$, composed of
wild type individuals with fitness 1 and mutants with fitness $r$.
Individuals are chosen at random (with rate $\mu$ for WT and $r\mu$
for mutant) to reproduce. When an individual reproduces, another is
chosen at random to die, therefore keeping the population at constant
size. The probabilities per unit of time for the mutants to increase
($W^{+})$ or decrease ($W^{-}$) their number by one individual is
(see appendix for a general explanation of stochastic modeling) 
\begin{eqnarray}
W^{+}(n) & = & r\mu(N-n)n/N\label{eq:altruistplus}\\
W^{-}(n) & = & \mu\left(N-n\right)n/N\label{eq:altruistsminus}
\end{eqnarray}
and their mean field approximation leads to the deterministic Fisher
equation (\ref{eq:detrm}). However, this is a probabilistic process
: the number of mutants can fall to zero (extinction) or $N$ (fixation)
with finite probability and if it does so, the system remains in this
state. One of the most fundamental concepts of evolutionary dynamics
is precisely the \emph{fixation probability}, \emph{i.e. }the probability
that a mutant spreads and takes over the whole community(\cite{Patwa2008a}).
In the framework of the Moran model the fixation probability is \cite{Moran1962,Houchmandzadeh2010}

\begin{equation}
\pi_{f}=\frac{1-r^{-N_{0}}}{1-r^{-N}}\label{eq:Moran0}
\end{equation}
where $N_{0}$ is the original number of mutants. For small selection
pressure $Ns\ll1$ where $s=r-1$, the fixation probability $\pi_{f}$
of a mutant appearing at one copy can be approximated by 
\begin{equation}
\pi_{f}\approx\frac{1}{N}+\frac{s}{2}\label{eq:fix1}
\end{equation}
The fixation probability is composed of two terms : even in the absence
of selection, the population will become homogenic via a process known
as genetic drift; in the neutral case, all individuals at generation
zero have an equal probability $1/N$ of of taking over the whole
community (being fixed). When a beneficial mutant is present, the
fixation probability of its carrier is increased by the relative excess
fitness. Note that  genetic drift is at the heart of the neutral theory
of biodiversity discussed in the previous section. 

The Fisher-Wright-Moran model is the most fundamental model of population
genetics, displaying the importance of genetic drift. We can complement
it to take into account the effect of altruistic individuals, without
adding any hidden benefits. The most notable effect of ``common good''
production is the increase in the carrying capacity of the habitat,
which benefits everybody regardless of its genotype (altruistic or
selfish). 
\begin{figure}
\begin{centering}
\includegraphics[width=0.8\columnwidth]{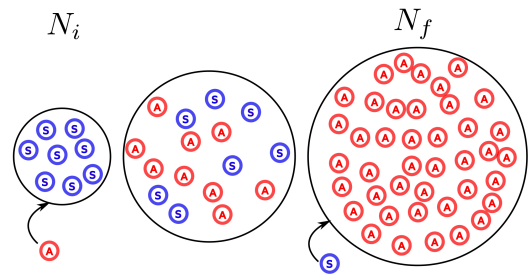}
\par\end{centering}

\caption{The neutral effect of common good production : the carrying capacity
$N$ of the habitat depends on the number of altruists present, ranging
from a minimum $N_{i}$ when only selfish individuals are present
to a maximum $N_{f}$. In this case, the fixation probability of one
A introduced into a community of S can be higher than the fixation
probability of one S introduced into a community of A.\label{fig:VariableN}}

\end{figure}
 Let us suppose that the carrying capacity is $N_{i}$ when only selfish
individuals are present and $N_{f}$ when only altruistic individuals
are present ($N_{i}<N_{f}$) and has an intermediate value when the
community is a mixture of both genotypes, with the carrying capacity
an increasing function of the number of altruistic individuals $n$
(figure \ref{fig:VariableN}). Let us suppose now that selfish individuals
have fitness $1$ and altruistic one have fitness $r<1$ and set $s=1-r$
as the cost of altruism. In a deterministic model, altruists will
\emph{always} lose to selfish ones\cite{Houchmandzadeh2012a}. When
taking into account the stochastic nature of this process, the answer
can be different. As I stressed above, the quantity of interest in
the stochastic process is the fixation probability. Let us compare
the fixation probability $\pi^{A}$ of one altruistic mutant introduced
into a community of selfish individuals to the fixation probability
$\pi^{S}$ of one selfish mutant introduced into a community of altruistic
individuals(figure \ref{fig:VariableN}). A back-of-the-envelop computation,
according to eq.(\ref{eq:fix1}) gives :
\[
\pi^{A}=\frac{1}{N_{i}}-\frac{s}{2}\,\,\,;\,\,\,\pi^{S}=\frac{1}{N_{f}}+\frac{s}{2}
\]
We see that even though selfish individuals have higher fitness, we
can nave $\pi^{A}>\pi^{S}$ if 
\[
s<\frac{1}{N_{i}}-\frac{1}{N_{f}}
\]
Alternatively, by setting $\Delta N=N_{f}-N_{i}$ and $\bar{N}=\sqrt{N_{i}N_{f}}$,
the above criteria can be written in terms of selection pressure 
\begin{equation}
\bar{N}s<\Delta N/\bar{N}\label{eq:selectionpressure}
\end{equation}
which means that if the selection pressure against the altruists is
smaller than the relative change in the carrying capacity, then altruists
\emph{win}, even though they have a smaller fitness. 

The above discussion is an over simplification in order to give the
general idea of the genetic drift favoring altruists. At each time
step, the community is formed of a number of $n$ of altruists and
$m$ of selfish individuals, and the total size of the community $N=n+m$
itself is a function of altruists numbers. The fixation probability
thus have to be computed by taking into account the full stochastic
behavior off both populations. The fixation probabilities can be computed
from the backward Kolmogorov (BK) equation and its diffusion approximation\cite{Ewens2004}.
This procedure was used by Kimura\cite{Kimura1962} to deduce the
fixation probability of the Fisher-Wright classical model of population
genetics. In this model the population size $N$ is kept constant
and the size of one population can be deduced from the size of the
other : $m=N-n$. The Kimura equation is thus concerned with only
the frequency of one allele. A similar approach can be used to compute
the fixation probability of the model presented above, although now
the BK equation has to track the frequency of both populations at
the same time, as the total population is not constant anymore. The
resulting equation, although much more complicated, leads to the result
(\ref{eq:selectionpressure}) given above\cite{Houchmandzadeh2012a}.

One could think that natural communities are composed of large number
of individuals, so even for small costs $s$, the criteria (\ref{eq:selectionpressure})
is violated. This argument however is not correct because populations
are geographically structured: individuals can be replaced only by
their neighbors, so the effective populations entering into  expression
(\ref{eq:selectionpressure}) are indeed much smaller than the total
size of the community. In fact, at small migration rate, the altruistic
advantage is amplified: at constant and uniform migration rate, the
number of migrants is proportional to the local population size; therefore
places with high carrying capacity (composed of altruists) send out
more migrants than places with a lower carrying capacity (composed
of selfish). This amplification mechanism can be computed at small
migration rates and it can be shown that large, geographically structured
population are indeed \emph{immune} to invasion by selfish individuals
($\pi^{S}=0$)\cite{Houchmandzadeh2012a}. 

Let us again stress that this simple advantage of altruists is a pure
effect of the discreteness of life which cannot exist if living organisms
were part of a continuum. I do not claim that kin or group selection
do not exist or are irrelevant, but there is an inherent advantage
in producing a common good that, when this trait increases the carrying
capacity of the habitat and benefits everybody. It may not overcome
the cost associated with this behavior in some living ecosystems and
then other more elaborate schemes have to be considered, but before
resorting to these ``hidden advantage'' theories, one should subtract
the contribution of discreteness and the increase in the carrying
capacity.

\section{Conclusion.}

There are many other biological systems where the discreteness of
underlying processes have come to the forefront. The most notable
example is noise driven chemical reactions taking place inside living
cells giving rise to  non-genetic individuality and which has been
thoroughly investigated during the last decade\cite{Davidson2008}.
The message I intended to carry through the three examples reviewed
in this paper is that, as in many other areas of science, the discrete
nature of life has important consequences which have been all too
often neglected. The main reason of this neglect may be the counterintuitive
nature of these consequences : a drop of ink in water tends to dilute
and it is not evident that by adding neutral reproduction, the ink
should reverse its course and \emph{concentrate}. I hope however that
this very fundamental and important aspect of life will become more
a part of the general culture of scientists.

\section{appendix : Stochastic modeling.}

A \emph{deterministic} behavior is perfectly predictable : knowing
that at time $t$ a system (for example a projectile) is in a state
$x$ (from example its position), we know its position $x'+x$ at
time $t+t'$. This knowledge is modeled by a function $x'=f(x,t,t')$
and completely characterizes the temporal dynamics of this particular
system. For many law it is enough to know the function $f$ for very
short (infinitesimal) time increment $t'$ which is then noted $dt$.
The evolution of the system for the small increment $x'=dx$ is then
often obtained as $dx=f(x,t)dt$ and called a differential equation.
It is enough to add these short increments to determine the state
of the system for any long time. 

A \emph{stochastic }behavior is only partially predictable : knowing
the system is in a state $x$ at time $t$, we cannot predict its
state at a later time $t+t'$; but only give a probability $P(x')$
that it will be at $x+x'$ at this time. The function $P(x')$ can
be experimentally measured by making a large number of measurement:
let $N$ similar systems at time $t$ in the state $x$ and measure
their state $x+x'$ at time $t+t'$, then make a normalized histogram
of all this measurements which is the function $P(x'|x,t,t')$. Again,
if we know this function for very short time $t'$, we can know the
probability function for longer times by summing up the short term
evolution. This is for example how meteorology works : knowing the
evolution of probabilities on time scales of seconds, we can predict
that for example there is 70\% chance that tomorrow will be rainy
and 30\% that it will be sunny. 

For many stochastic systems (called Markov chains), it is enough to
know the temporal evolution of probabilities for infinitesimal time
increments $t'=dt$ written as
\[
P(x'|x,t,dt)=W(x'|x,t)dt
\]
where the function $W$ is called the transition rate. Equations (\ref{eq:transition rates}),
(\ref{eq:metacom1},\ref{eq:metacom2}) and (\ref{eq:altruistplus},\ref{eq:altruistsminus})
are examples of such transition rates. $W^{\pm}(n)$ are short hand
for $W(\pm1|n)$, \emph{i.e.} the probability (per unit of time) that
the system (here number of individuals of a given type) will be in
state $n+1$ at time $t+dt$, knowing that it is in state $n$ at
time $t$. The knowledge of these short time transition rates allows
for the determination of probabilities at long time, through an evolution
equation called the Master equation when the states are discrete. 

It not always easy to solve a Master equation. A crude approximation,
called mean field, is to write a \emph{deterministic} equation for
the evolution of the \emph{mean} increment $\left\langle dx\right\rangle $
during a short time $dt$, in the form of 
\[
\left\langle dx\right\rangle =f(x,t)dt
\]
and then use deterministic procedure to predict the \emph{mean} state
of the system at long times. The function $f(x,t)$ is readily obtained
from the transition rates $W(x'|x,t)$. Mean field approximation can
be useful or misleading, depending on the stochastic nature of the
system. 

\bibliographystyle{plainnat}
\bibliography{/home/bahram/0Papers/Bibtex/mutspace,/home/bahram/0Papers/Bibtex/Almora_2012,/home/bahram/0Papers/Bibtex/Doctoral_Evolution,/home/bahram/0Papers/Altruism_2011/bibli_altruism,/home/bahram/0Papers/Bibtex/microbiology,/home/bahram/0Papers/PUBLISHED/sto_clust2007/clustering}

\begin{thebibliography}{58}
\expandafter\ifx\csname natexlab\endcsname\relax\def\natexlab#1{#1}\fi
\expandafter\ifx\csname bibnamefont\endcsname\relax
  \def\bibnamefont#1{#1}\fi
\expandafter\ifx\csname bibfnamefont\endcsname\relax
  \def\bibfnamefont#1{#1}\fi
\expandafter\ifx\csname citenamefont\endcsname\relax
  \def\citenamefont#1{#1}\fi
\expandafter\ifx\csname url\endcsname\relax
  \def\url#1{\texttt{#1}}\fi
\expandafter\ifx\csname urlprefix\endcsname\relax\def\urlprefix{URL }\fi
\providecommand{\bibinfo}[2]{#2}
\providecommand{\eprint}[2][]{\url{#2}}

\bibitem[{\citenamefont{Boomsma} \emph{et~al.}(2011)\citenamefont{Boomsma,
  Beekman, Cornwallis, Griffin, Holman, Hughes, Keller, Oldroyd, and
  Ratnieks}}]{Boomsma2011}
\bibinfo{author}{\bibnamefont{Boomsma}, \bibfnamefont{J.~J.}},
  \bibinfo{author}{\bibfnamefont{M.}~\bibnamefont{Beekman}},
  \bibinfo{author}{\bibfnamefont{C.~K.} \bibnamefont{Cornwallis}},
  \bibinfo{author}{\bibfnamefont{A.~S.} \bibnamefont{Griffin}},
  \bibinfo{author}{\bibfnamefont{L.}~\bibnamefont{Holman}},
  \bibinfo{author}{\bibfnamefont{W.~O.~H.} \bibnamefont{Hughes}},
  \bibinfo{author}{\bibfnamefont{L.}~\bibnamefont{Keller}},
  \bibinfo{author}{\bibfnamefont{B.~P.} \bibnamefont{Oldroyd}}, and
  \bibinfo{author}{\bibfnamefont{F.~L.~W.} \bibnamefont{Ratnieks}},
  \bibinfo{year}{2011}, \bibinfo{journal}{Nature}
  \textbf{\bibinfo{volume}{471}}, \bibinfo{pages}{E4}.

\bibitem[{\citenamefont{Butlin} \emph{et~al.}(2008)\citenamefont{Butlin,
  Galindo, and Grahame}}]{Butlin2008}
\bibinfo{author}{\bibnamefont{Butlin}, \bibfnamefont{R.~K.}},
  \bibinfo{author}{\bibfnamefont{J.}~\bibnamefont{Galindo}}, and
  \bibinfo{author}{\bibfnamefont{J.~W.} \bibnamefont{Grahame}},
  \bibinfo{year}{2008}, \bibinfo{journal}{Philosophical transactions of the
  Royal Society of London. Series B, Biological sciences}
  \textbf{\bibinfo{volume}{363}}, \bibinfo{pages}{2997}.

\bibitem[{\citenamefont{Condit} \emph{et~al.}(2002)\citenamefont{Condit,
  Pitman, Leigh, Chave, Terborgh, Foster, N\'{u}\~{n}ez, Aguilar, Valencia,
  Villa, Muller-Landau, Losos} \emph{et~al.}}]{Condit2002}
\bibinfo{author}{\bibnamefont{Condit}, \bibfnamefont{R.}},
  \bibinfo{author}{\bibfnamefont{N.}~\bibnamefont{Pitman}},
  \bibinfo{author}{\bibfnamefont{E.~G.} \bibnamefont{Leigh}},
  \bibinfo{author}{\bibfnamefont{J.}~\bibnamefont{Chave}},
  \bibinfo{author}{\bibfnamefont{J.}~\bibnamefont{Terborgh}},
  \bibinfo{author}{\bibfnamefont{R.~B.} \bibnamefont{Foster}},
  \bibinfo{author}{\bibfnamefont{P.}~\bibnamefont{N\'{u}\~{n}ez}},
  \bibinfo{author}{\bibfnamefont{S.}~\bibnamefont{Aguilar}},
  \bibinfo{author}{\bibfnamefont{R.}~\bibnamefont{Valencia}},
  \bibinfo{author}{\bibfnamefont{G.}~\bibnamefont{Villa}},
  \bibinfo{author}{\bibfnamefont{H.~C.} \bibnamefont{Muller-Landau}},
  \bibinfo{author}{\bibfnamefont{E.}~\bibnamefont{Losos}}, \emph{et~al.},
  \bibinfo{year}{2002}, \bibinfo{journal}{Science (New York, N.Y.)}
  \textbf{\bibinfo{volume}{295}}, \bibinfo{pages}{666}.

\bibitem[{\citenamefont{Davidson and Surette}(2008)}]{Davidson2008}
\bibinfo{author}{\bibnamefont{Davidson}, \bibfnamefont{C.~J.}}, and
  \bibinfo{author}{\bibfnamefont{M.~G.} \bibnamefont{Surette}},
  \bibinfo{year}{2008}, \bibinfo{journal}{Annual review of genetics}
  \textbf{\bibinfo{volume}{42}}, \bibinfo{pages}{253}.

\bibitem[{\citenamefont{Dugatkin}(2006)}]{Dugatkin2006}
\bibinfo{author}{\bibnamefont{Dugatkin}, \bibfnamefont{L.}},
  \bibinfo{year}{2006}, \emph{\bibinfo{title}{The Altruism Equation}}
  (\bibinfo{publisher}{Princeton University Press}).

\bibitem[{\citenamefont{Durrett and Levin}(1994)}]{Durrett1994}
\bibinfo{author}{\bibnamefont{Durrett}, \bibfnamefont{R.}}, and
  \bibinfo{author}{\bibfnamefont{S.}~\bibnamefont{Levin}},
  \bibinfo{year}{1994}, \bibinfo{journal}{Theoretical Population Biology}
  \textbf{\bibinfo{volume}{46}}, \bibinfo{pages}{363}.

\bibitem[{\citenamefont{Etienne}(2005)}]{Etienne2005}
\bibinfo{author}{\bibnamefont{Etienne}, \bibfnamefont{R.~S.}},
  \bibinfo{year}{2005}, \bibinfo{journal}{Ecology Letters}
  \textbf{\bibinfo{volume}{8}}, \bibinfo{pages}{253}.

\bibitem[{\citenamefont{Ewens}(2004)}]{Ewens2004}
\bibinfo{author}{\bibnamefont{Ewens}, \bibfnamefont{W.~J.}},
  \bibinfo{year}{2004}, \emph{\bibinfo{title}{Mathematical Population
  Genetics}} (\bibinfo{publisher}{Springer-Verlag}).

\bibitem[{\citenamefont{Fisher}(1999)}]{Fisher1999}
\bibinfo{author}{\bibnamefont{Fisher}, \bibfnamefont{R.}},
  \bibinfo{year}{1999}, \emph{\bibinfo{title}{The genetical theory of natural
  selection, a complete variorum edition}} (\bibinfo{publisher}{Oxford
  University Press}).

\bibitem[{\citenamefont{Foster} \emph{et~al.}(2004)\citenamefont{Foster,
  Shaulsky, Strassmann, Queller, and Thompson}}]{Foster2004}
\bibinfo{author}{\bibnamefont{Foster}, \bibfnamefont{K.~R.}},
  \bibinfo{author}{\bibfnamefont{G.}~\bibnamefont{Shaulsky}},
  \bibinfo{author}{\bibfnamefont{J.~E.} \bibnamefont{Strassmann}},
  \bibinfo{author}{\bibfnamefont{D.~C.} \bibnamefont{Queller}}, and
  \bibinfo{author}{\bibfnamefont{C.~R.~L.} \bibnamefont{Thompson}},
  \bibinfo{year}{2004}, \bibinfo{journal}{Nature}
  \textbf{\bibinfo{volume}{431}}, \bibinfo{pages}{693}.

\bibitem[{\citenamefont{Gore} \emph{et~al.}(2009)\citenamefont{Gore, Youk, and
  van Oudenaarden}}]{Gore2009}
\bibinfo{author}{\bibnamefont{Gore}, \bibfnamefont{J.}},
  \bibinfo{author}{\bibfnamefont{H.}~\bibnamefont{Youk}}, and
  \bibinfo{author}{\bibfnamefont{A.}~\bibnamefont{van Oudenaarden}},
  \bibinfo{year}{2009}, \bibinfo{journal}{Nature}
  \textbf{\bibinfo{volume}{459}}, \bibinfo{pages}{253}.

\bibitem[{\citenamefont{Gould and Lewontin}(1979)}]{gould1979}
\bibinfo{author}{\bibnamefont{Gould}, \bibfnamefont{S.}}, and
  \bibinfo{author}{\bibfnamefont{R.}~\bibnamefont{Lewontin}},
  \bibinfo{year}{1979}, \bibinfo{journal}{Proc. R. Soc. Lond. B}
  \textbf{\bibinfo{volume}{205}}, \bibinfo{pages}{581}.

\bibitem[{\citenamefont{Hamilton}(1964{\natexlab{a}})}]{Hamilton1964a}
\bibinfo{author}{\bibnamefont{Hamilton}, \bibfnamefont{W.~D.}},
  \bibinfo{year}{1964}{\natexlab{a}}, \bibinfo{journal}{J Theor Biol}
  \textbf{\bibinfo{volume}{7}}, \bibinfo{pages}{1}.

\bibitem[{\citenamefont{Hamilton}(1964{\natexlab{b}})}]{Hamilton1964b}
\bibinfo{author}{\bibnamefont{Hamilton}, \bibfnamefont{W.~D.}},
  \bibinfo{year}{1964}{\natexlab{b}}, \bibinfo{journal}{J Theor Biol}
  \textbf{\bibinfo{volume}{7}}, \bibinfo{pages}{17}.

\bibitem[{\citenamefont{Hanski and Gaggiotti}(2004)}]{Hanski2004}
\bibinfo{editor}{\bibnamefont{Hanski}, \bibfnamefont{I.}}, and
  \bibinfo{editor}{\bibfnamefont{O.}~\bibnamefont{Gaggiotti}} (eds.),
  \bibinfo{year}{2004}, \emph{\bibinfo{title}{Ecology, Genetics and Evolution
  of Metapopulations}} (\bibinfo{publisher}{Academic Press}).

\bibitem[{\citenamefont{Hanski and Gilpin}(1997)}]{Hanski1997}
\bibinfo{editor}{\bibnamefont{Hanski}, \bibfnamefont{I.}}, and
  \bibinfo{editor}{\bibfnamefont{M.}~\bibnamefont{Gilpin}} (eds.),
  \bibinfo{year}{1997}, \emph{\bibinfo{title}{{Metapopulation Biology}}}
  (\bibinfo{publisher}{Academic Press Inc}).

\bibitem[{\citenamefont{Hardin}(1960)}]{Hardin1960}
\bibinfo{author}{\bibnamefont{Hardin}, \bibfnamefont{G.}},
  \bibinfo{year}{1960}, \bibinfo{journal}{science}
  \textbf{\bibinfo{volume}{131}}, \bibinfo{pages}{1292}.

\bibitem[{\citenamefont{Houchmandzadeh}(2002)}]{Houchmandzadeh2002}
\bibinfo{author}{\bibnamefont{Houchmandzadeh}, \bibfnamefont{B.}},
  \bibinfo{year}{2002}, \bibinfo{journal}{Phys Rev E}
  \textbf{\bibinfo{volume}{66}}, \bibinfo{pages}{052902}.

\bibitem[{\citenamefont{Houchmandzadeh}(2008)}]{Houchmandzadeh2008}
\bibinfo{author}{\bibnamefont{Houchmandzadeh}, \bibfnamefont{B.}},
  \bibinfo{year}{2008}, \bibinfo{journal}{Phys Rev Lett}
  \textbf{\bibinfo{volume}{101}}(\bibinfo{number}{7}), \bibinfo{pages}{078103}.

\bibitem[{\citenamefont{Houchmandzadeh}(2009)}]{Houchmandzadeh2009a}
\bibinfo{author}{\bibnamefont{Houchmandzadeh}, \bibfnamefont{B.}},
  \bibinfo{year}{2009}, \bibinfo{journal}{Physical Review E}
  \textbf{\bibinfo{volume}{80}}, \bibinfo{pages}{051920}.

\bibitem[{\citenamefont{Houchmandzadeh and
  Mihalcescu}(2011)}]{Houchmandzadeh2011a}
\bibinfo{author}{\bibnamefont{Houchmandzadeh}, \bibfnamefont{B.}}, and
  \bibinfo{author}{\bibfnamefont{I.}~\bibnamefont{Mihalcescu}},
  \bibinfo{year}{2011}, \bibinfo{journal}{Europhysics News}
  \textbf{\bibinfo{volume}{42}}, \bibinfo{pages}{36}.

\bibitem[{\citenamefont{Houchmandzadeh and Vallade}(2003)}]{Houchmandzadeh2003}
\bibinfo{author}{\bibnamefont{Houchmandzadeh}, \bibfnamefont{B.}}, and
  \bibinfo{author}{\bibfnamefont{M.}~\bibnamefont{Vallade}},
  \bibinfo{year}{2003}, \bibinfo{journal}{Phys Rev E}
  \textbf{\bibinfo{volume}{68}}, \bibinfo{pages}{061912}.

\bibitem[{\citenamefont{Houchmandzadeh and Vallade}(2010)}]{Houchmandzadeh2010}
\bibinfo{author}{\bibnamefont{Houchmandzadeh}, \bibfnamefont{B.}}, and
  \bibinfo{author}{\bibfnamefont{M.}~\bibnamefont{Vallade}},
  \bibinfo{year}{2010}, \bibinfo{journal}{Phys Rev E}
  \textbf{\bibinfo{volume}{82}}, \bibinfo{pages}{051913}.

\bibitem[{\citenamefont{Houchmandzadeh and
  Vallade}(2012)}]{Houchmandzadeh2012a}
\bibinfo{author}{\bibnamefont{Houchmandzadeh}, \bibfnamefont{B.}}, and
  \bibinfo{author}{\bibfnamefont{M.}~\bibnamefont{Vallade}},
  \bibinfo{year}{2012}, \bibinfo{journal}{BMC evolutionary biology}
  \textbf{\bibinfo{volume}{12}}, \bibinfo{pages}{61}.

\bibitem[{\citenamefont{Hu} \emph{et~al.}(2006)\citenamefont{Hu, He, and
  Hubbell}}]{Hu2006}
\bibinfo{author}{\bibnamefont{Hu}, \bibfnamefont{X.-S.}},
  \bibinfo{author}{\bibfnamefont{F.}~\bibnamefont{He}}, and
  \bibinfo{author}{\bibfnamefont{S.~P.} \bibnamefont{Hubbell}},
  \bibinfo{year}{2006}, \bibinfo{journal}{Oikos}
  \textbf{\bibinfo{volume}{113}}, \bibinfo{pages}{548}.

\bibitem[{\citenamefont{Hubbel}(2001)}]{Hubbel2001}
\bibinfo{author}{\bibnamefont{Hubbel}, \bibfnamefont{S.~P.}},
  \bibinfo{year}{2001}, \emph{\bibinfo{title}{The unified neutral theory of
  Biodiversity and Biogeography.}} (\bibinfo{publisher}{Princeton University
  Press}).

\bibitem[{\citenamefont{Jabot and Chave}(2011)}]{Jabot2011}
\bibinfo{author}{\bibnamefont{Jabot}, \bibfnamefont{F.}}, and
  \bibinfo{author}{\bibfnamefont{J.}~\bibnamefont{Chave}},
  \bibinfo{year}{2011}, \bibinfo{journal}{The American naturalist}
  \textbf{\bibinfo{volume}{178}}, \bibinfo{pages}{E37}.

\bibitem[{\citenamefont{Kessin}(2001)}]{Kessin2001}
\bibinfo{author}{\bibnamefont{Kessin}, \bibfnamefont{R.~H.}},
  \bibinfo{year}{2001}, \emph{\bibinfo{title}{Dictyostelium: Evolution, Cell
  Biology and the Development of Multicellularity.}}
  (\bibinfo{publisher}{Cambridge Univ. Press, Cambridge}).

\bibitem[{\citenamefont{Kimura}(1962)}]{Kimura1962}
\bibinfo{author}{\bibnamefont{Kimura}, \bibfnamefont{M.}},
  \bibinfo{year}{1962}, \bibinfo{journal}{Genetics}
  \textbf{\bibinfo{volume}{47}}, \bibinfo{pages}{713}.

\bibitem[{\citenamefont{Kimura}(1985)}]{Kimura1985}
\bibinfo{author}{\bibnamefont{Kimura}, \bibfnamefont{M.}},
  \bibinfo{year}{1985}, \emph{\bibinfo{title}{{The Neutral Theory of Molecular
  Evolution}}}, volume \bibinfo{volume}{1985}.

\bibitem[{\citenamefont{Lewontin}(1970)}]{Lewontin1970}
\bibinfo{author}{\bibnamefont{Lewontin}, \bibfnamefont{R.~C.}},
  \bibinfo{year}{1970}, \bibinfo{journal}{Annual Review of Ecology and
  Systematics} \textbf{\bibinfo{volume}{1}}, \bibinfo{pages}{1}.

\bibitem[{\citenamefont{Lomolino}(2006)}]{Lo}
\bibinfo{author}{\bibnamefont{Lomolino}, \bibfnamefont{M.~V.}},
  \bibinfo{year}{2006}, \emph{\bibinfo{title}{{Biogeography.}}}
  (\bibinfo{publisher}{Sinauer Associates Inc.,U.S.}).

\bibitem[{\citenamefont{MacArthur and MacArthur}(1967)}]{MacArthur1967}
\bibinfo{author}{\bibnamefont{MacArthur}, \bibfnamefont{R.~H.}}, and
  \bibinfo{author}{\bibfnamefont{R.~H.} \bibnamefont{MacArthur}},
  \bibinfo{year}{1967}, \emph{\bibinfo{title}{{The Theory of Island
  Biogeography}}} (\bibinfo{publisher}{[[Princeton University Press]]},
  \bibinfo{address}{Princeton, N.J.}).

\bibitem[{\citenamefont{MacArthur and Wilson}(1963)}]{MacArthur1963}
\bibinfo{author}{\bibnamefont{MacArthur}, \bibfnamefont{R.~H.}}, and
  \bibinfo{author}{\bibfnamefont{E.~O.} \bibnamefont{Wilson}},
  \bibinfo{year}{1963}, \bibinfo{journal}{Evolution}
  \textbf{\bibinfo{volume}{17}}, \bibinfo{pages}{373}.

\bibitem[{\citenamefont{Mal\'{e}cot}(1948)}]{Malecot1948}
\bibinfo{author}{\bibnamefont{Mal\'{e}cot}, \bibfnamefont{G.}},
  \bibinfo{year}{1948}, \emph{\bibinfo{title}{{Les math\'{e}matiques de
  l'h\'{e}r\'{e}dit\'{e}}}} (\bibinfo{publisher}{Masson, Paris}).

\bibitem[{\citenamefont{Mayr}(1942)}]{Mayr1942}
\bibinfo{author}{\bibnamefont{Mayr}, \bibfnamefont{E.}}, \bibinfo{year}{1942},
  \emph{\bibinfo{title}{{Systematics and the Origin of Species: From the
  Viewpoint of a Zoologist}}} (\bibinfo{publisher}{Harvard University Press}).

\bibitem[{\citenamefont{McGill} \emph{et~al.}(2006)\citenamefont{McGill,
  Maurer, and Weiser}}]{McGill2006}
\bibinfo{author}{\bibnamefont{McGill}, \bibfnamefont{B.~J.}},
  \bibinfo{author}{\bibfnamefont{B.~A.} \bibnamefont{Maurer}}, and
  \bibinfo{author}{\bibfnamefont{M.~D.} \bibnamefont{Weiser}},
  \bibinfo{year}{2006}, \bibinfo{journal}{Ecology}
  \textbf{\bibinfo{volume}{87}}, \bibinfo{pages}{1411}.

\bibitem[{\citenamefont{Moran}(1962)}]{Moran1962}
\bibinfo{author}{\bibnamefont{Moran}, \bibfnamefont{P.}}, \bibinfo{year}{1962},
  \emph{\bibinfo{title}{The Statistical processes of of evolutionary theory}}
  (\bibinfo{publisher}{Oxford University Press}).

\bibitem[{\citenamefont{Novick and Weiner}(1957)}]{Novick1957}
\bibinfo{author}{\bibnamefont{Novick}, \bibfnamefont{A.}}, and
  \bibinfo{author}{\bibfnamefont{M.}~\bibnamefont{Weiner}},
  \bibinfo{year}{1957}, \bibinfo{journal}{PNAS} \textbf{\bibinfo{volume}{43}},
  \bibinfo{pages}{553}.

\bibitem[{\citenamefont{Nowak}(2006{\natexlab{a}})}]{Nowak}
\bibinfo{author}{\bibnamefont{Nowak}, \bibfnamefont{M.~A.}},
  \bibinfo{year}{2006}{\natexlab{a}}, \emph{\bibinfo{title}{{Evolutionary
  Dynamics: Exploring the Equations of Life}}} (\bibinfo{publisher}{The Belknap
  Press}).

\bibitem[{\citenamefont{Nowak}(2006{\natexlab{b}})}]{Nowak2006}
\bibinfo{author}{\bibnamefont{Nowak}, \bibfnamefont{M.~A.}},
  \bibinfo{year}{2006}{\natexlab{b}}, \bibinfo{journal}{Science}
  \textbf{\bibinfo{volume}{314}}, \bibinfo{pages}{1560}.

\bibitem[{\citenamefont{Nowak} \emph{et~al.}(2010)\citenamefont{Nowak, Tarnita,
  and Wilson}}]{Nowak2010}
\bibinfo{author}{\bibnamefont{Nowak}, \bibfnamefont{M.~A.}},
  \bibinfo{author}{\bibfnamefont{C.~E.} \bibnamefont{Tarnita}}, and
  \bibinfo{author}{\bibfnamefont{E.~O.} \bibnamefont{Wilson}},
  \bibinfo{year}{2010}, \bibinfo{journal}{Nature}
  \textbf{\bibinfo{volume}{466}}, \bibinfo{pages}{1057}.

\bibitem[{\citenamefont{Okubo and Levin}(2002)}]{Okubo2002}
\bibinfo{author}{\bibnamefont{Okubo}, \bibfnamefont{A.}}, and
  \bibinfo{author}{\bibfnamefont{S.~A.} \bibnamefont{Levin}},
  \bibinfo{year}{2002}, \emph{\bibinfo{title}{{Diffusion and Ecological
  Problems, Modern Perspectives, Second Edition}}}
  (\bibinfo{publisher}{Springer}).

\bibitem[{\citenamefont{O'Malley}(2007)}]{O'Malley2007}
\bibinfo{author}{\bibnamefont{O'Malley}, \bibfnamefont{M.~a.}},
  \bibinfo{year}{2007}, \bibinfo{journal}{Nature reviews. Microbiology}
  \textbf{\bibinfo{volume}{5}}, \bibinfo{pages}{647}.

\bibitem[{\citenamefont{Patwa and Wahl}(2008)}]{Patwa2008a}
\bibinfo{author}{\bibnamefont{Patwa}, \bibfnamefont{Z.}}, and
  \bibinfo{author}{\bibfnamefont{L.~M.} \bibnamefont{Wahl}},
  \bibinfo{year}{2008}, \bibinfo{journal}{J R Soc Interface}
  \textbf{\bibinfo{volume}{5}}, \bibinfo{pages}{1279}.

\bibitem[{\citenamefont{Price}(1970)}]{Price1970}
\bibinfo{author}{\bibnamefont{Price}, \bibfnamefont{G.~R.}},
  \bibinfo{year}{1970}, \bibinfo{journal}{Nature}
  \textbf{\bibinfo{volume}{227}}, \bibinfo{pages}{520}.

\bibitem[{\citenamefont{Rosindell} \emph{et~al.}(2011)\citenamefont{Rosindell,
  Hubbell, and Etienne}}]{Rosindell2011}
\bibinfo{author}{\bibnamefont{Rosindell}, \bibfnamefont{J.}},
  \bibinfo{author}{\bibfnamefont{S.~P.} \bibnamefont{Hubbell}}, and
  \bibinfo{author}{\bibfnamefont{R.~S.} \bibnamefont{Etienne}},
  \bibinfo{year}{2011}, \bibinfo{journal}{Trends in ecology \& evolution}
  \textbf{\bibinfo{volume}{26}}, \bibinfo{pages}{340}.

\bibitem[{\citenamefont{Strassmann}
  \emph{et~al.}(2011)\citenamefont{Strassmann, Page, Robinson, and
  Seeley}}]{Strassmann2011}
\bibinfo{author}{\bibnamefont{Strassmann}, \bibfnamefont{J.~E.}},
  \bibinfo{author}{\bibfnamefont{R.~E.} \bibnamefont{Page}},
  \bibinfo{author}{\bibfnamefont{G.~E.} \bibnamefont{Robinson}}, and
  \bibinfo{author}{\bibfnamefont{T.~D.} \bibnamefont{Seeley}},
  \bibinfo{year}{2011}, \bibinfo{journal}{Nature}
  \textbf{\bibinfo{volume}{471}}, \bibinfo{pages}{E5}.

\bibitem[{\citenamefont{Taylor} \emph{et~al.}(1978)\citenamefont{Taylor,
  Woiwod, and Perry}}]{Taylor1978}
\bibinfo{author}{\bibnamefont{Taylor}, \bibfnamefont{L.~R.}},
  \bibinfo{author}{\bibfnamefont{I.~P.} \bibnamefont{Woiwod}}, and
  \bibinfo{author}{\bibfnamefont{J.~N.} \bibnamefont{Perry}},
  \bibinfo{year}{1978}, \bibinfo{journal}{J. Anim. Ecol.}
  \textbf{\bibinfo{volume}{47}}, \bibinfo{pages}{383}.

\bibitem[{\citenamefont{Traulsen} \emph{et~al.}(2005)\citenamefont{Traulsen,
  Sengupta, and Nowak}}]{Traulsen2005}
\bibinfo{author}{\bibnamefont{Traulsen}, \bibfnamefont{A.}},
  \bibinfo{author}{\bibfnamefont{A.~M.} \bibnamefont{Sengupta}}, and
  \bibinfo{author}{\bibfnamefont{M.~A.} \bibnamefont{Nowak}},
  \bibinfo{year}{2005}, \bibinfo{journal}{J Theor Biol}
  \textbf{\bibinfo{volume}{235}}, \bibinfo{pages}{393}.

\bibitem[{\citenamefont{Vallade and Houchmandzadeh}(2003)}]{Vallade2003a}
\bibinfo{author}{\bibnamefont{Vallade}, \bibfnamefont{M.}}, and
  \bibinfo{author}{\bibfnamefont{B.}~\bibnamefont{Houchmandzadeh}},
  \bibinfo{year}{2003}, \bibinfo{journal}{Phys Rev E Stat Nonlin Soft Matter
  Phys} \textbf{\bibinfo{volume}{68}}, \bibinfo{pages}{61902}.

\bibitem[{\citenamefont{Vallade and Houchmandzadeh}(2006)}]{Vallade2006}
\bibinfo{author}{\bibnamefont{Vallade}, \bibfnamefont{M.}}, and
  \bibinfo{author}{\bibfnamefont{B.}~\bibnamefont{Houchmandzadeh}},
  \bibinfo{year}{2006}, \bibinfo{journal}{Phys Rev E Stat Nonlin Soft Matter
  Phys} \textbf{\bibinfo{volume}{74}}, \bibinfo{pages}{51914}.

\bibitem[{\citenamefont{Visick and Ruby}(2006)}]{Visick2006}
\bibinfo{author}{\bibnamefont{Visick}, \bibfnamefont{K.~L.}}, and
  \bibinfo{author}{\bibfnamefont{E.~G.} \bibnamefont{Ruby}},
  \bibinfo{year}{2006}, \bibinfo{journal}{Current opinion in microbiology}
  \textbf{\bibinfo{volume}{9}}, \bibinfo{pages}{632}.

\bibitem[{\citenamefont{Volkov} \emph{et~al.}(2003)\citenamefont{Volkov,
  Banavar, Hubbell, and Maritan}}]{Volkov2003}
\bibinfo{author}{\bibnamefont{Volkov}, \bibfnamefont{I.}},
  \bibinfo{author}{\bibfnamefont{J.~R.} \bibnamefont{Banavar}},
  \bibinfo{author}{\bibfnamefont{S.~P.} \bibnamefont{Hubbell}}, and
  \bibinfo{author}{\bibfnamefont{A.}~\bibnamefont{Maritan}},
  \bibinfo{year}{2003}, \bibinfo{journal}{Nature}
  \textbf{\bibinfo{volume}{424}}, \bibinfo{pages}{1035}.

\bibitem[{\citenamefont{West and Buckling}(2003)}]{West2003}
\bibinfo{author}{\bibnamefont{West}, \bibfnamefont{S.~A.}}, and
  \bibinfo{author}{\bibfnamefont{A.}~\bibnamefont{Buckling}},
  \bibinfo{year}{2003}, \bibinfo{journal}{Proc Biol Sci}
  \textbf{\bibinfo{volume}{270}}, \bibinfo{pages}{37}.

\bibitem[{\citenamefont{Williams}(1966)}]{Williams1966}
\bibinfo{author}{\bibnamefont{Williams}, \bibfnamefont{G.~C.}},
  \bibinfo{year}{1966}, \emph{\bibinfo{title}{{Adaptation and Natural
  Selection: A Critique of Some Current Evolutionary Thought}}}
  (\bibinfo{publisher}{Princeton University Press}).

\bibitem[{\citenamefont{Williamson}
  \emph{et~al.}(2008)\citenamefont{Williamson, Gaston, and
  Lonsdale}}]{Williamson2008}
\bibinfo{author}{\bibnamefont{Williamson}, \bibfnamefont{M.}},
  \bibinfo{author}{\bibfnamefont{K.~J.} \bibnamefont{Gaston}}, and
  \bibinfo{author}{\bibfnamefont{W.~M.} \bibnamefont{Lonsdale}},
  \bibinfo{year}{2008}, \bibinfo{journal}{Journal of Biogeography}
  \textbf{\bibinfo{volume}{28}}, \bibinfo{pages}{827}.

\bibitem[{\citenamefont{Young} \emph{et~al.}(2001)\citenamefont{Young, Roberts,
  and Stuhne}}]{Young2001}
\bibinfo{author}{\bibnamefont{Young}, \bibfnamefont{W.~R.}},
  \bibinfo{author}{\bibfnamefont{A.~J.} \bibnamefont{Roberts}}, and
  \bibinfo{author}{\bibfnamefont{G.}~\bibnamefont{Stuhne}},
  \bibinfo{year}{2001}, \bibinfo{journal}{Nature}
  \textbf{\bibinfo{volume}{412}}, \bibinfo{pages}{328}.

\end{thebibliography}

\end{document}